\documentclass[aps,twocolumn,prl,showpacs,floatfix,amsmath,amssymb,superscriptaddress]{revtex4-1}
\usepackage{graphicx}
\usepackage{bm}
\usepackage{amsmath,amssymb}
\usepackage{hyperref}
\usepackage{color}
\usepackage{natbib} 
\usepackage{braket,float,soul}
\usepackage{comment}
\begin{document}

\title{Topology in the space-time scaling limit of quantum dynamics}

\author{Lorenzo  Rossi}
\email{lorenzo.rossi@polito.it}
\affiliation{Dipartimento di Scienza Applicata e Tecnologia, Politecnico di Torino, 10129 Torino, Italy}

\author{Jan Carl Budich}
\affiliation{Institute of Theoretical Physics, Technische Universit\"at Dresden and
W\"urzburg-Dresden Cluster of Excellence ct.qmat, 01062 Dresden, Germany}

\author{Fabrizio Dolcini}
\affiliation{Dipartimento di Scienza Applicata e Tecnologia, Politecnico di Torino, 10129 Torino, Italy}

 \begin{abstract}
We investigate the role of topology in the space-time scaling limit of quantum quench dynamics, where both time and system size tend to infinity at a constant ratio.  There, while the standard topological characterization relying on local unitary transformations becomes ill defined, we show how a different dynamical notion of topology naturally arises through a dynamical winding number encoding the linear response of the Berry phase to a magnetic flux.
 Specifically, we find that the presence of a locally invisible constant magnetic flux is revealed by a dynamical staircase behavior of the Berry phase, whose topologically quantized plateaus characterize the space-time scaling limit of a quenched Rice-Mele model.  These jumps in the Berry phase are also shown to be related to the interband elements of the DC current operator. We outline possible experimental platforms for observing the predicted phenomena in finite systems. 
\end{abstract}

\maketitle

Topology has become a cornerstone for understanding and distinguishing phases of matter \cite{Kane_RMP_2010,Ryu_RMP_2016,Wen_RMP_2017}. 
While initially this approach has mostly been used to unravel topological properties of low temperature systems \cite{vonKlitzing_PRL_1980,Thouless_PRL_1982,Haldane_PRL_1988,Kane_PRL_2005bis,Zhang_Science_2006,Molenkamp_Science_2007}, recent advances in experimentally controlling the quantum dynamics of atomic many-particle states \cite{Goldman_NatPhys_2016,Cooper_RMP_2019} have triggered the study of topological features far from equilibrium.
In particular, within the paradigmatic quantum quench protocol\cite{Silva_RMP_2011}, new dynamical topological invariants, which are predicted to characterize the change in topology of the quenched Hamiltonian\cite{Dora_PRB_2015,Budich_PRB_2016,Balatsky_PRL_2016,Zhai_PRL_2017,Chen_PRB_2018,Slager_PRL_2020} have been observed\cite{Weitenberg_NatPhys_2018,Weitenberg_NatComm_2019}, and the dynamical robustness of topological features has been addressed, both theoretically\cite{Gurarie_PRB_2013,Sacramento_PRE_2014,Gurarie_PRL_2014,Fazio_PRB_2014,Cooper_PRL_2015,Fazio_PRB_2016,Toniolo_PRB_2018,Cooper_PRL_2018,Cooper_PRB_2019} and experimentally\cite{Spielman_PRL_2022}. Since topological phases may be defined as equivalence classes under local unitary transformations \cite{Chen2010}, bulk topological properties of a quantum state cannot dynamically change during coherent time evolution generated by a local Hamiltonian\cite{Chen2010, Gurarie_PRB_2013,Sacramento_PRE_2014,Cooper_PRL_2015,Chen_PRB_2018,Cooper_RMP_2019}.
Notwithstanding these fundamental constraints, symmetry protected topological invariants can be fragile, if the underlying symmetries are dynamically broken \cite{Cooper_PRL_2018,Cooper_PRB_2019,Spielman_PRL_2022}. 
In addition, topological invariants are typically defined in the thermodynamic limit (TL), while all experiments deal with finite systems. Hence, the conventional topological characterization is meaningful only for time scales such that $t\ll L/v^f$, where $L$ measures the system size, and~$v^f$ is a characteristic band velocity of the post quench Hamiltonian \cite{Cooper_RMP_2019}.
At later times, since an extensively long unitary time evolution is no longer a local transformation, standard topological properties are expected to become ill defined \cite{Cooper_RMP_2019} and previous works  on quantum quenches  have thus mostly focused on the $t\ll L/v^f$ regime.
 Alternatively, the opposite regime $t\gg L/v^f$ has been addressed in the context of adiabatic state preparation, where the finite size is harnessed to adiabatically connect different equilibrium topological phases \cite{Rigol_PRA_2017, Pollmann_PRB_2017, Vishwanath_PRB_2017, Lukin_Nature_2019, Budich_PRL_2020}.  

\begin{figure}
\centering
\includegraphics[width=\linewidth]{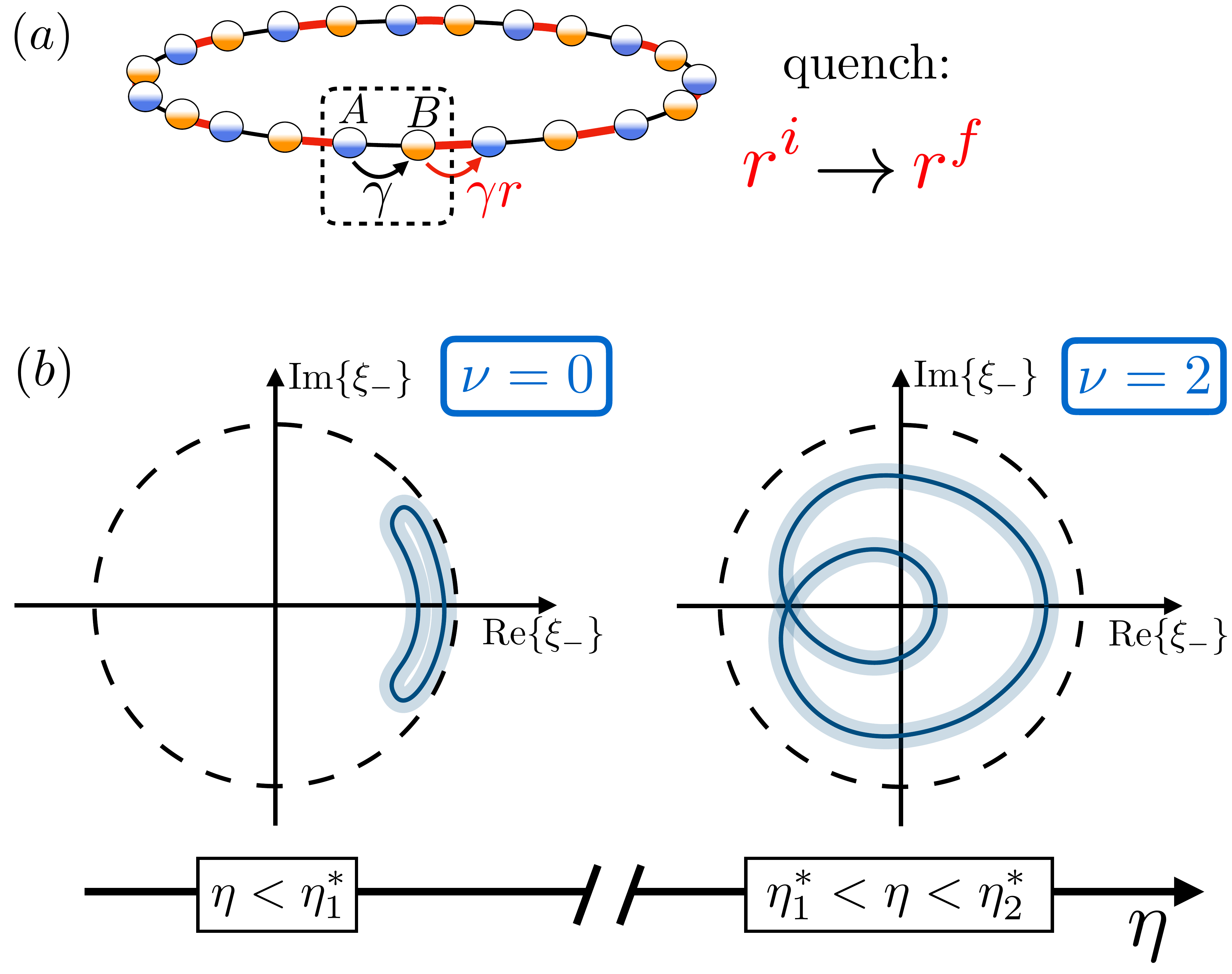}
\caption{\label{Fig1} (a) Illustration of a Rice-Mele lattice model on a ring [see Eq.~(\ref{RiceMele})],  subject to a quench by a sudden variation of the intercell hopping amplitude  $\gamma r^i\rightarrow \gamma r^f $. (b) Schematic representation of the closed loops formed in the complex plane by the Bloch state overlap $\xi_-$ [see Eq.~(\ref{xi_-})] as a function of quasi-momentum $k$. Non-trivial loops (right panel) may form in the STSL regime at  at critical values $\eta^*_m$ [see Eq.~(\ref{eta^*})] of the ratio $\eta =2\pi t/L$. Solid lines represent the $0$-th order contribution $\xi_-^{(0)}(k,\eta)$ [see Eq.~(\ref{xi_-^{(0)}})], shallow halos visualize the sub-leading contribution  $2\pi\xi_-^{(1)}(k,\eta)/L$. When $\eta<\eta_1^*$ (left panel) the winding number $\nu$ vanishes, while for $\eta_1^*<\eta<\eta_2^*$ (right panel) $\nu=2$. The dashed line depicts the unit circle as a guide to the eye.}
\end{figure}

\begin{figure}
\centering
\includegraphics[width=\linewidth]{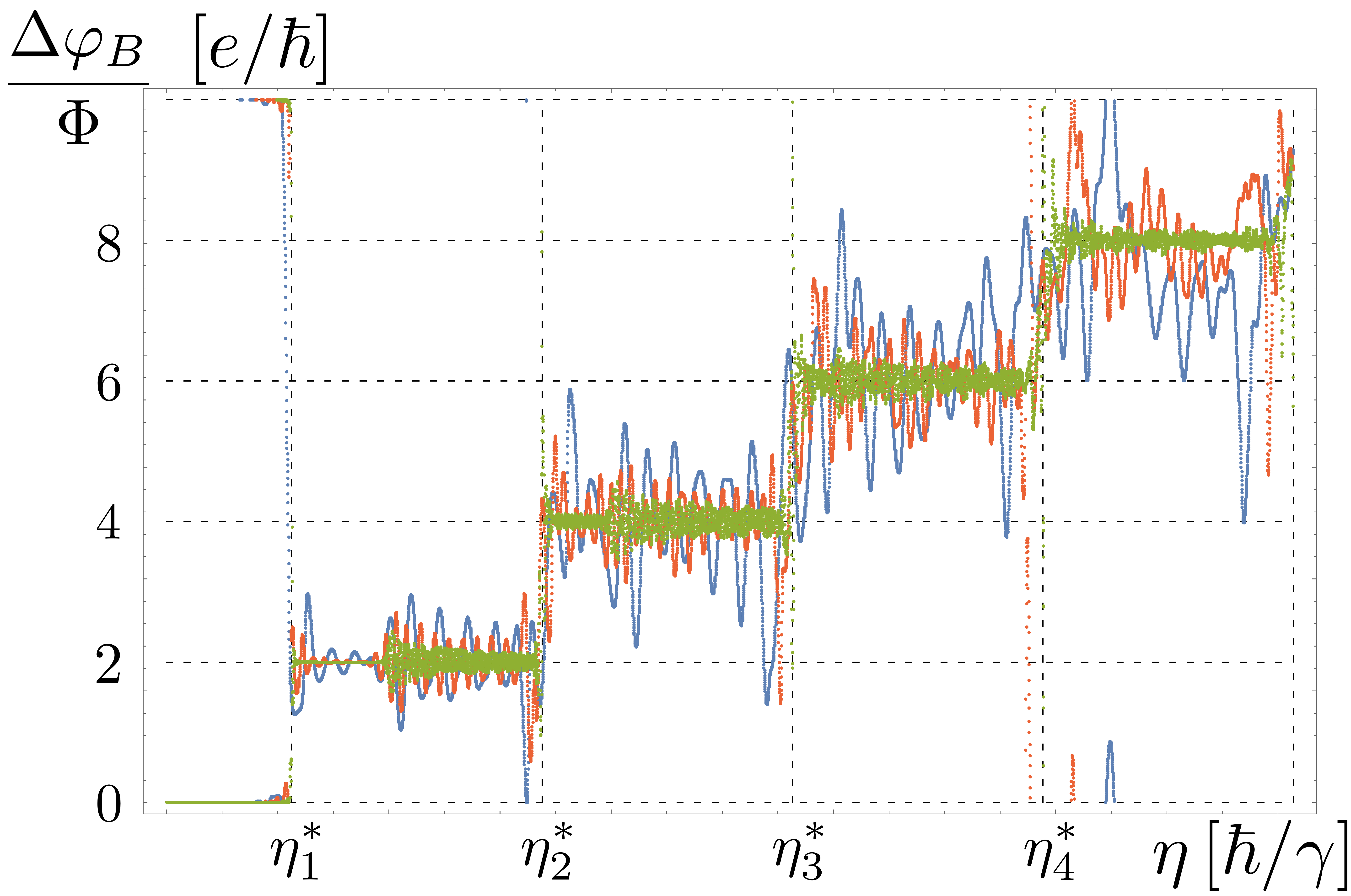}
\caption{\label{Fig2}  The linear response $\Delta \varphi_B/\Phi$ of the Berry phase to an applied magnetic flux is plotted, in units of $e/\hbar=2\pi/\Phi_0$,  as a function of $\eta=2\pi t/{L}$ in the STSL regime, after quantum quenches in finite Rice-Mele lattices with PBC [see~Eq.~(\ref{RiceMele})]. In all quenches, the energy scale $\gamma$ is fixed to a constant value throughout the entire protocol, together with the ratio of the staggered potential $u=0.1$. The ratio $r$ of the staggered hopping amplitudes is instead quenched from $r^i=0.5$ to $r^f=2$ while the magnetic flux, when present, is constant and equal to $\Phi/\Phi_0=1/10$. System sizes are ${L}=40$ (blue), ${L}=80$ (red), and ${L}=400$ (green).  The plateaus at $\nu=2$, $4$, $6$, and $8$ are clearly visible already for ${L}=40$, they do not depend on system size and abruptly change at critical ratios $\eta_m^*$. The fluctuations are instead system size dependent and are suppressed with increasing $L$.} 
\end{figure}

In the present work we propose to investigate a different out of equilibrium regime, namely the quench dynamics in the space-time scaling limit (STSL), where \textit{both} time and system size tend to infinity while their ratio $\eta= 2\pi t/L$ is kept constant, and we  show how a different dynamical topological invariant $\nu(\eta)$ naturally arises (see Fig.~\ref{Fig1} for an illustration with a quenched Rice-Mele model\cite{Rice_PRL_1982,Palyi_Book_2016}).  
To understand its physical implications, we analyze the effect of a constant magnetic flux $\Phi$ threading a one dimensional (1D) system with periodic boundary conditions (PBC). Remarkably, 
while $\Phi$, as a global property, remains invisible in the quench dynamics for sub-extensive times, in the STSL the Berry phase \cite{Berry_ProcRoySocLond_1984,Simon_PRL_1983,Palyi_Book_2016} is found to dynamically acquire a staircase behavior (see Fig.~\ref{Fig2}), whose plateau values are topologically quantized as $2\pi \nu \Phi/\Phi_0$, where $\Phi_0=h/e$ is the flux quantum. Since the limits $t\rightarrow + \infty$ and $L\rightarrow + \infty$ do not commute, these properties are unique to the STSL regime and cannot be obtained by applying the long time limit to formulae derived in the standard TL. Yet, we demonstrate that clear signatures of our predictions can be observed in finite systems of moderate size that are within reach of present day quantum simulators.

For definiteness, we consider a sudden quench in a system of non-interacting spinless fermions hopping in a 1D bipartite lattice with PBC, and we assume the Hamiltonian to be traceless.
We measure lengths in units of the lattice spacing $a$, so that the length $L$ of the system coincides with the number of cells.
Thanks to translation invariance we can write the initial/final realizations in reciprocal space as $H^{i/f}=\sum_k c^\dagger(k)[\mathbf{d}^{i/f}(k) \cdot \boldsymbol{\sigma}] c(k)$.
Here $k\in 2\pi n/{L}$ is a dimensionless quasi-momentum, where $n\in\{-\lfloor {L}/2\rfloor, \ldots, \lfloor ({L}-1)/2\rfloor \} $, while $\boldsymbol{\sigma}$ is the three dimensional vector of Pauli matrices and $c^\dagger(k)=(c^\dagger_A(k),c^\dagger_B(k))$ is a spinor of fermionic operators, which create spinless fermions with quasi-momentum $k$ in sublattice $A/B$.
All the information about the specific Hamiltonian realizations is thus encoded in the $k$-dependent three dimensional vectors $\mathbf{d}^{i/f}(k)$. 
In particular,  the initial/final spectra are given by $\epsilon_\pm^{i/f}(k)=\pm|\mathbf{d}^{i/f}(k)|$.
Moreover, the time evolved many-particle state can be easily reconstructed out of the single particle time dependent Bloch spinors $|u_\pm(k,t)\rangle=e^{-i[\mathbf{d}^f(k)\cdot\boldsymbol{\sigma}]t/\hbar}|u_\pm^i(k)\rangle$, where $|u_\pm^i(k)\rangle$ are the Bloch single particle eigenstates of $H^i$.

We assume $H^i$ to have a finite band gap, initialize the system in its half filled insulating ground state, and follow the time evolution of the Berry phase in its discretized formulation, appropriate for finite system sizes $\varphi_B(t,{L})=\sum_k \arg \xi_-(k,t,{L})~$\cite{Suzuki_JPhysSocJapan_2005}, where
\begin{equation}\label{xi_-}
    \xi_-(k,t,{L})=\langle u_-(k+\delta k,t)|u_-(k,t)\rangle
\end{equation}
and $\delta k =2\pi/{L}$. 
As in the standard continuous formulation, the discrete Berry phase is gauge invariant under $|u_-(k,t)\rangle \rightarrow |u_-^\lambda(k,t)\rangle =e^{i\lambda(k)}|u_-(k,t)\rangle$ and takes quantized values, equal to either $0$ or $\pi$, when charge conjugation symmetry is present\cite{Ryu_RMP_2016,Palyi_Book_2016,SM}.
Moreover, in the usual TL, i.e. ${L}\rightarrow + \infty$ while $t\in \mathbb{R}$, it is straightforward to realize that $\xi_-(k,t,{L})=1+iA_B(k,t)\delta k  + O({L}^{-2})$, where $A_B(k,t)=\langle u_-(k,t)|i\partial_k |u_-(k,t)\rangle$ is the time dependent Berry connection, and the standard result $\varphi_B(t)=\int_{-\pi}^\pi dk \, A_B(k,t)$ is recovered \cite{Palyi_Book_2016}.

However, in the STSL, when $t,{L}\rightarrow+\infty$ with fixed $\eta=\delta k \, t = 2\pi t/{L} \in \mathbb{R}$, the function $\xi_-(k,t,{L})$ may develop a non-trivial dependence on $k$ and $\eta$ already to zeroth order in the $1/{L}$ expansion.
Indeed one can write $\xi_-(k,t,{L})=\xi_-^{(0)}(k,\eta) + \xi_-^{(1)}(k,\eta,t)\delta k + O({L}^{-2})$ where\cite{SM}
\begin{equation}\label{xi_-^{(0)}}
\xi_-^{(0)}(k,\eta)=\cos [ v^f(k) \eta] -i\, \mathcal{C}(k) \sin [ v^f(k) \eta]  \quad . 
\end{equation}
Here $\mathcal{C}(k)=\hat{\mathbf{d}}^i(k) \cdot \hat{\mathbf{d}}^f(k)$ is the cosine of the $k$-dependent angle between the initial and final unit vectors, while $v^f(k)=\partial_k \epsilon_+^f(k)/\hbar$ is the post quench band velocity. 
Then it is straightforward to derive $|\xi_-^{(0)}(k,\eta)|=\sqrt{1-\{ \mathcal{S}(k) \sin[v^f(k)\eta] \}^2}$, where $\mathcal{S}^2(k)=1-\mathcal{C}^2(k)$, and we notice that, if $\mathcal{C}(k)=0$ is satisfied by some $k^*$, Eq.(\ref{xi_-^{(0)}}) vanishes at equally spaced critical ratios
\begin{equation}\label{eta^*}
    \eta_m^*  =  \left(\frac{\pi}{2}+ (m-1)\pi \right)\frac{1}{v^f(k^*)} , \quad m \in \mathbf{N}_+ \quad .
\end{equation}
Some comments are in order. 
In the limit $\eta \rightarrow 0$ one has $\xi_-^{(0)}(k,\eta)\rightarrow 1+O({L}^{-1})$ and the standard TL result is recovered.
Moreover, at finite $\eta$, Eq.(\ref{xi_-^{(0)}}) is reminiscent of the $k$-dependent contribution to the Loschmidt amplitude, appearing in the context of dynamical quantum phase transitions (DQPT)\cite{Heyl_PRL_2013,Dora_PRB_2015,Budich_PRB_2016}. 
Similarly, the condition $\mathcal{C}(k^*)=0$ leading to a vanishing $\xi_-^{(0)}$ in Eq.(\ref{xi_-^{(0)}}) is formally equivalent to the requirement for observing DQPT\cite{Dora_PRB_2015}.
However, we emphasize that, while the $k$ dependent contribution to the Loshmidt amplitude stems from the overlap between the initial and the time evolved Bloch spinor at the same $k$, the quantity studied here, Eq.(\ref{xi_-}), is the overlap between Bloch spinors that are {\it both} time evolved and that are computed at {\it different} quasi-momenta, namely $k$ and $k + \delta k$. 
It is precisely such a tiny deviation that yields to Eq.(\ref{xi_-^{(0)}}) at $t\sim {L}/v^f$.
Thus, while DQPT occur at \textit{finite} times in a TL system, Eq.(\ref{xi_-^{(0)}}) vanishes at {\it extensive} critical times $t^*_m=\eta^*_m L/2\pi$, with  $\eta^*_m$ given by Eq.(\ref{eta^*}).

We now start to investigate the topological features unique to the STSL regime, i.e. where $\eta$ takes finite values  even for arbitrarily large systems. Far away from its critical values, by treating $\eta$ as a parameter, we can define $\alpha^{(0)}(k;\eta)=\arg \xi_-^{(0)}(k,\eta)$. The function $k \mapsto \alpha^{(0)}(k;\eta)$ from a circle to a circle naturally leads to the definition of a dynamical winding number $\nu(\eta)\in \mathbb{Z}$ through $\alpha^{(0)}(k;\eta)=\tilde\alpha^{(0)}(k;\eta)+ k \, \nu(\eta)$, where $\tilde\alpha^{(0)}(k;\eta)$ is a $\mathbb{R}$-valued smooth periodic function.
Remarkably, by contrast to the conventional equilibrium framework \cite{Ryu_RMP_2016}, this dynamical winding number does not require any symmetry to be properly defined.
We can then write the Berry phase in the STSL regime as $\varphi_B(\eta)=\varphi_B^{(0)}(\eta)+\varphi_B^{(1)}(\eta)$, where 
\begin{equation}\label{varphi_B^{(0)}}
    \varphi_B^{(0)}(\eta)=\frac{{L}}{2\pi} \int_{-\pi}^\pi dk \, [\tilde\alpha^{(0)}(k;\eta)+ k \, \nu(\eta)] \quad ,
\end{equation}
while $\varphi_B^{(1)}(\eta)$ is analogous to the usual integral of the Berry connection\cite{SM}.
Thus, let us focus on the consequences of the new contribution stemming from a non-trivial $\xi_-^{(0)}$.
A priori, $\varphi_B^{(0)}(\eta)$ is of order $L$ and, given that the Berry phase is defined $\mod 2\pi$, the zeroth order would produce a Berry phase that wildly fluctuates with time.
Nonetheless, if $\mathbf{d}^i(k)$ and $\mathbf{d}^f(k)$ have the same parity under $k \leftrightarrow -k$, then $\alpha^{(0)}(k;\eta)$ becomes an odd function of $k$ and the integral in Eq.(\ref{varphi_B^{(0)}}) vanishes identically.
This condition physically corresponds to a quench that does not generate any stationary current\cite{Rossi_PRB_2022}.
However, if we now assume that a finite and constant magnetic flux $\Phi$ is present throughout the entire quench dynamics, the quasi-momenta get shifted according to $k\rightarrow k + \phi$, where $\phi=\frac{2\pi}{{L}}\frac{\Phi}{\Phi_0}$.
This shift does not affect the integral of the odd periodic part $\tilde\alpha^{(0)}(k;\eta)$, which remains vanishing.
However, although~$\phi$ is infinitesimal for large ${L}$, the shift yields a finite contribution proportional to $\nu(\eta)$, thanks to the factor~${L}$ in Eq.(\ref{varphi_B^{(0)}}). 
We thus end up with
\begin{equation}\label{varphi_B^{(0)}(Phi)}
    \varphi_B^{(0)}(\eta;\Phi) = 2\pi \nu(\eta) \Phi/\Phi_0 + O({L}^{-1}) \quad .
\end{equation}
We can therefore conclude that, in the STSL, the Berry phase develops a non-trivial zeroth order contribution which induces a quantized response to an applied magnetic flux and the quantization is encoded in the dynamical topological invariant $\nu(\eta)$.
 In this respect, $\nu$ plays a role analogous to the Chern number in the integer quantum Hall effect \cite{vonKlitzing_PRL_1980,Thouless_PRL_1982}: while the latter uniquely defines the linear Hall response to an applied electric field, the former encodes the linear response of the Berry phase to an applied magnetic flux.
However, while the various plateaus in the Hall conductance identifies different equilibrium topological phases as a function of the chemical potential, the winding number $\nu(\eta)$ topologically characterizes an out of equilibrium state and is thus a function of time. 
Note that, since $\nu(\eta)$ can change only at the critical ratios $\eta^*_m$ in Eq.(\ref{eta^*}), the topological invariant is stable for {\it extensive} time windows $\Delta t={L} / 2 v^f(k^*)$. This means that the system undergoes a new kind of dynamical topological phase transition, where a well defined topological invariant suddenly changes at the extensive critical  times $t^*_m$.
Moreover, since the quantized response does not depend on system size, it is remarkable to notice that even a fraction of the elementary flux quantum may yield a detectable signature in the coherent dynamics of a macroscopic quantum system.

After the above general derivations, we now choose a specific setup to illustrate our results. 
We consider a sudden quench of the hopping amplitudes in the Rice-Mele model, which is defined by
\begin{equation}\label{RiceMele}
\mathbf{d}(k)=\gamma(1+r\cos k, r\sin k, u)
\end{equation}
and depicted in Fig.\ref{Fig1}(a). Here $\gamma$ is the reference energy scale, $r$ is the ratio between intercell and intracell hopping, and $u$ is the ratio between the staggered potential on $A$ and $B$ sublattice, breaking charge conjugation and chiral symmetry.
We choose a quench such that $\mathcal{C}(k)$ vanishes for some $k^*$, a condition that,  for   a given $H^f$, is fulfilled by a vast class of initial states. Here we quench from $r^i=0.5$ to $r^f=2$ while keeping $u=-0.1$ constant.
It is then straightforward to show that $\nu(\eta)$, which has to be zero for $\eta=0$, increases by two at each critical ratio $\eta^*_m$.
Such increase by two units can be easily understood if one recognizes that the condition $\mathcal{C}(k)=0$ is satisfied by two quasi-momenta $\{k^*_1,k^*_2\}$ which, because of symmetry, are related by $k_1^*=-k_2^*$ and are thus associated to the same critical ratios $\eta^*_m$.
Concurrently, the closed loop traced by $\xi_-^{(0)}(k,\eta)$ in the complex plane as a function of $k$ touches the origin twice at the critical ratios and the winding increases by two.
Far away from $\eta^*_m$ the winding of $\xi_-^{(0)}(k,\eta)$ is instead a robust topological invariant. 
Moreover, it coincides with the winding of the whole overlap function Eq.(\ref{xi_-}), since the first order contribution $\xi_-^{(1)}(k,\eta) \delta k$ is suppressed by a factor ${L}^{-1}$ and it cannot destroy the robustness of the invariant.
A comparison between the loops traced by $\xi_-(k,\eta)$ for $\eta<\eta_1^*$ and $\eta_1^*<\eta<\eta_2^*$ is schematically depicted in Fig.\ref{Fig1}(b), where the solid lines denote the finite contribution given by $\xi_-^{(0)}(k,\eta)$, while the shallow halos around them account for the ${L}^{-1}$ contribution carried by $\xi_-^{(1)}(k,\eta) \delta k$.

We can now fully appreciate the interplay between a finite dynamical winding number and a constant magnetic flux.
In Fig.\ref{Fig2}, we plot the $\eta$-dependent response of the Berry phase  to an applied magnetic flux, namely $\Delta \varphi_B(\eta)/\Phi$ where $\Delta \varphi_B(\eta)=\varphi_B(\eta;\Phi \ne 0)-\varphi_B(\eta;\Phi = 0)$, for the above specified quench in a finite Rice-Mele lattice.
We compute the same quantity for different system sizes, while keeping the non-zero value of the magnetic flux always equal to $\Phi/\Phi_0=1/10$, and we display the values of $\Delta \varphi_B(\eta)/\Phi$ in units of the universal constants $e/\hbar=2\pi/\Phi_0$. 
 
Increasing ${L}$ at constant $\eta$, hence going towards the STSL regime, a staircase profile becomes more and more pronounced. 
The critical ratios at which the jumps occur are given by Eq.(\ref{eta^*}) while the heights of the different plateaus are encoded in Eq.(\ref{varphi_B^{(0)}(Phi)}).
The reason is straightforward:
The contribution to the Berry phase given by $\Delta \varphi_B^{(1)}(\eta)$ amounts to bounded fluctuations with zero average, which are produced by the slight mismatch between $k$ and $k+\phi$ and are suppressed in the STSL. 
The contribution carried by $\Delta \varphi_B^{(0)}(\eta)$ instead corresponds to rigid shifts of $4\pi \Phi/\Phi_0$ each time a critical ratio is reached, independent of system size.
In the proper STSL a sharp staircase profile is thus recovered.

We would like to elaborate on the differences between the TL and the STSL in terms of the Berry phase, the Wannier wavefunctions, and the particle current density.
In the standard TL ($\eta\rightarrow 0$), the many-particle insulating state can be built out of a Slater determinant of exponentially localized Wannier functions\cite{Marzari_PRB_1997}. 
Because a vector potential can be always gauged away for such wavefunctions\cite{Kohn_PR_1964}, a constant magnetic flux cannot lead to observable signatures.
At the same time, the time derivative of the Berry phase is linked, even out of equilibrium, to the particle current density\cite{Cooper_PRL_2018}. 
In contrast, in the STSL regime, the localization length of the Wannier functions becomes comparable to system size\cite{footnote_Wannier}, with a twofold implication. On the one hand, the magnetic flux can no longer be gauged away and can lead to observable signatures, such as the staircase profile depicted in Fig.\ref{Fig2}.
On the other hand, the jumps of the Berry phase at the critical ratios $\eta_m^*$ are not associated to a physical current. Instead one can show that\cite{SM}
\begin{eqnarray}
\frac{d}{d\eta} \varphi_B^{(0)}(\eta) &=& \frac{{L}}{2\pi} \int_{-\pi}^\pi \! dk \, \langle u_-^i(k) | \mathcal{J}_{DC}^f(k) | u_-^i(k) \rangle \label{d varphi}\\
&&\hspace{-40pt}+ \frac{{L}}{2\pi} \int_{-\pi}^\pi \! dk \, \text{Re} \left\{ \frac{\chi_-^{(0)}(k,\eta)}{\xi_-^{(0)}(k,\eta)} \langle u_+^i(k) | \mathcal{J}_{DC}^f(k) | u_-^i(k) \rangle \right\}  \nonumber 
\end{eqnarray}
where $\mathcal{J}_{dc}^f(k)=v^f(k) \, \hat{\mathbf{d}}^f(k) \cdot \boldsymbol{\sigma}$ is the component of the particle current operator that commutes with the post quench Hamiltonian and describes a DC current, while $\chi_-^{(0)}(k,\eta)=\langle u_-(k+\delta k,t) | u_+(k,t) \rangle+O({L}^{-1})$.
In the interesting case in which the Berry phase develops a staircase profile, the first integral, which is the expectation value of the DC current and it is the only contribution appearing in the long time limit of a TL system, is vanishing due to symmetry.
The jumps are instead produced by the additional contribution in the second line of Eq.(\ref{d varphi}), which is absent in the standard TL. 
Such integral does not correspond to the expectation value of a particle current and it rather involves the inter-band elements of the DC current operator.

In summary, we have shown that intriguing topological features arise in the STSL regime after a quantum quench, when both time and system size are sent to infinity while keeping their ratio finite.
In particular, we have rigorously defined a  dynamical winding number $\nu(\eta)$, which characterizes the many-particle state of a 1D two-band model in the STSL regime, see Fig.\ref{Fig1}(b).
Notably, its definition does not rely on any specific symmetry, at variance with the customary equilibrium setting.
We have shown that the dynamical winding number physically encodes the linear response of the Berry phase to an applied magnetic flux, which thus exhibits a staircase behavior as a function of $\eta$, see Fig.\ref{Fig2}. The plateaus are quantized in units of $e/\hbar$ and the jumps between them occur at the well defined critical times given by Eq.(\ref{eta^*}).
It is also worth mentioning that this phenomenon can be observed with state of the art experimental techniques.
The long coherence time of ultracold atoms in optical lattices\cite{Schmiedmayer_AnnRevCondMatPhys_2015} may also allow one to approach the STSL regime experimentally in finite systems.
Moreover, given the possibility to generate artificial gauge fields\cite{Goldman_RepProgPhys_2014} and reconstruct the time dependent Berry phase through quantum state tomography techniques\cite{Weitenberg_NatPhys_2018,Weitenberg_NatComm_2019,Spielman_PRL_2022}, we expect experiments with ultracold atoms, similar to the one described in Ref.\cite{Spielman_PRL_2022}, to enable observing the onset of a staircase profile as depicted in Fig.\ref{Fig2}.
An alternative implementation could be based on quantum walks in photonic platforms where the present quench dynamics can be simulated and the time-dependent Berry phase  can be measured \cite{Cardano2017,Guo_LightScienceApplication_2020}.
Our work provides a starting point for investigating further topological properties unique to the STSL regime, including the study of higher dimensions with richer geometry of Bloch bands, and probing the robustness of the dynamical winding number $\nu$ to the breaking of translation invariance and its generalization in the presence of many-body interactions.

{\it Acknowledgements} L.R. acknowledges useful discussions with Luca Barbiero, Raphaël Saint-Jalm, and Ian Spielman. J.C.B. acknowledges financial support from the German Research Foundation (DFG) through the Collaborative Research Centre SFB 1143 (Project No. 247310070), the Cluster of Excellence ct.qmat (Project No. 390858490), and the DFG Project 419241108. F.D. acknowledges financial support from the Italian Centro Nazionale di Ricerca in High Performance Computing, Big Data and Quantum Computing, funded by European Union – NextGenerationEU (grant no. CN00000013).

\widetext
\clearpage

\setcounter{section}{0}
\setcounter{equation}{0}
\setcounter{figure}{0}
\setcounter{table}{0}
\setcounter{page}{1}
\makeatletter
\renewcommand{\theequation}{S\arabic{equation}}
\renewcommand{\thefigure}{S\arabic{figure}}
\renewcommand{\bibnumfmt}[1]{[S#1]}
\renewcommand{\citenumfont}[1]{S#1}

\section{TOPOLOGY IN THE SPACE-TIME SCALING LIMIT OF QUANTUM DYNAMICS \\ \vspace{5pt}
SUPPLEMENTAL MATERIAL}
In this Supplemental Material we summarize some useful results about quenches in two band models and we provide some details about the definition of the Berry phase for a finite size system, its evaluation in the space-time scaling limit, and its connection to the current operator. 

\subsection{Quench in a two band model}
In the quench protocol described in the Main Text, the single particle eigenstates $| u_\pm^i  (k) \rangle$ of the initial Hamiltonian $H^i=\sum_k c^\dagger(k) [\mathbf{d}^i(k)\cdot \boldsymbol\sigma] c(k)$, determined by the unit vector $\hat{\mathbf{d}}^i(k)=\mathbf{d}^i(k)/|\mathbf{d}^i(k)|$ through the eigenvalue problem $[\hat{\mathbf{d}}^i(k)\cdot \boldsymbol\sigma] | u_\pm^i  (k) \rangle=\pm | u_\pm^i  (k) \rangle$,  evolve according to the post-quench Hamiltonian $H^f=\sum_k c^\dagger(k) [\mathbf{d}^f(k)\cdot \boldsymbol\sigma] c(k)$ as $| u_\pm(k,t) \rangle  = \exp[-i [\mathbf{d}^f(k) \cdot \boldsymbol{\sigma}] t/\hbar  ]  | u_\pm^i  (k) \rangle$. In turn, the related projector $\rho_\pm(k,t)=| u_\pm(k,t)\rangle \langle u_\pm(k,t) |$  can always be written as $\rho_\pm(k,t)=   [ \sigma_0 \pm \hat{\mathbf{d}}(k,t)\cdot \boldsymbol{\sigma}  ]/2$, where $\sigma_0$ denotes the $2 \times 2$  identity matrix  and $\hat{\mathbf{d}}(k,t)$ is  a time-dependent unit vector  given by \cite{Chen_PRB_2018_SM}
\begin{eqnarray}
 \hat{\mathbf{d}}(k,t)= \mathbf{d}_\parallel(k)+\mathbf{d}_\bot(k) \cos[2|\mathbf{d}^f(k)|t/\hbar]+\mathbf{d}_\times(k)\sin[2|\mathbf{d}^f(k)|t/\hbar]\quad,\end{eqnarray}
where $\mathbf{d}_\parallel(k)= [ \hat{\mathbf{d}}^i(k) \cdot \hat{\mathbf{d}}^f(k) ]\hat{\mathbf{d}}^f(k)$, $\mathbf{d}_\bot(k) = \hat{\mathbf{d}}^i(k) - \mathbf{d}_\parallel(k)$ and $\mathbf{d}_\times(k) = -[ \hat{\mathbf{d}}^i(k) \times \hat{\mathbf{d}}^f(k) ]$, with $\hat{\mathbf{d}}^f(k)=\mathbf{d}^f(k)/|\mathbf{d}^f(k)|$.
Because the initial state is the many-body  ground state of the initial Hamiltonian $H^i$ at half filling, where the lower band $\epsilon_{-}^i(k)=-|\mathbf{d}^i(k)|$ is completely filled and the upper band $\epsilon_{+}^i(k)=+|\mathbf{d}^i(k)|$ is empty, the evolved many-body state can always be regarded to as the half filled ground state of a fictitious (two flat band) Hamiltonian $H(t)=\sum_k c^\dagger(k) [\hat{\mathbf{d}}(k,t)\cdot \boldsymbol\sigma] c(k)$, where time appear as a parameter.

\subsection{Discrete Berry phase}
On the basis of the previous Section, the Berry phase associated to the many body state of a quenched two band insulator can be formulated in terms of $|u_-(k,t)\rangle $ and $\hat{\mathbf{d}}(k,t)$.  
We shall now determine some of its general properties that hold independently of the time dependence. 
In order to lighten the notation, we are thus going to omit the $t$ variable, which will be restored later,  when time plays a major role.
Yet we will deal with both the lower band Berry phase $\varphi_{B-}$ and the upper band Berry phase $\varphi_{B+}$.
In the Main Text, the symbol $\varphi_B$ was used to denote $\varphi_{B-}$ while here we explicitly keep both band indexes to highlight the relations between $\varphi_{B-}$ and $\varphi_{B+}$.

When dealing with a finite system, the Berry phase has to be reformulated in terms of finite differences according to $\varphi_{B\pm}=\sum_{k\in BZ}\arg\xi_{\pm}\, \mod 2\pi$, where $\xi_{\pm}= \langle u_\pm(k+\delta k)|u_\pm(k) \rangle$, $BZ$ denotes the Brillouin Zone and $\delta k = 2\pi/L$~\cite{Suzuki_JPhysSocJapan_2005_SM}. 

We first note that in the continuum limit $L\rightarrow + \infty$, if $|u_\pm(k) \rangle$ is a smooth function of $k$ (as it is usually the case in the time independent framework and for $t\ll L/v^f$), we can approximate $\xi_{\pm} \approx 1+i\langle  u_\pm(k)|i\partial_k |u_\pm(k) \rangle \delta k$, implying $\arg \xi_{\pm} \approx \langle  u_\pm(k)|i\partial_k |u_\pm(k) \rangle \delta k$, and we recover the standard expression $\varphi_{B\pm}=\int dk A_{B\pm}(k)$, where $A_{B\pm}(k)=\langle u_\pm(k)|i\partial_k |u_\pm(k)\rangle$ is the Berry connection of the upper/lower band. 

Then we show that some general properties fulfilled by the Berry phase in the customary continuous formulation are preserved also in the present discrete  formulation \cite{Palyi_Book_2016_SM}. Indeed, despite $\xi_{\pm}$ at each $k$ is gauge dependent, the Berry phase is gauge invariant, as can be seen by rewriting $\varphi_{B\pm}$ as 
\begin{eqnarray}
    \varphi_{B\pm}&=&   \text{Im} \ln \prod_{k\in BZ} \langle u_\pm(k+\delta k)|u_\pm(k) \rangle   = \arg \text{tr} \prod_{k\in BZ} \rho_\pm(k)  \label{Eq: Berry phase explicit gauge ind.}
\end{eqnarray}
where $\text{tr}$ denotes the trace on  a two dimensional space and the projectors $\rho_\pm(k)$  are gauge invariant. 

Moreover, Eq.(\ref{Eq: Berry phase explicit gauge ind.}) also enables one to prove that $\varphi_{B-}+\varphi_{B+}=0 \mod 2\pi$. 
Indeed we observe that, for each $k$, the related projector  can be written as $\rho_\pm=  \beta^0(\pm\hat{\mathbf{d}}) \sigma_0 + \boldsymbol{\beta}(\pm\hat{\mathbf{d}})\cdot \boldsymbol{\sigma}$, where the real  parts $\beta^0_R,\boldsymbol{\beta}_R$ and the imaginary parts $\beta^0_I,\boldsymbol{\beta}_I$ 
of the $\beta$-coefficients are  functions of the unit vector  $\hat{\mathbf{d}}=\hat{\mathbf{d}}(k)$    satisfying the following parity relations
\begin{equation} \label{Eq: parity relations}
    \begin{cases} 
        \beta^0_R(\hat{\mathbf{d}})=+\beta^0_R(-\hat{\mathbf{d}}) \\
        \beta^0_I(\hat{\mathbf{d}})=-\beta^0_I(-\hat{\mathbf{d}}) \\
        \boldsymbol{\beta}_R(\hat{\mathbf{d}})=-\boldsymbol{\beta}_R(-\hat{\mathbf{d}}) \\
        \boldsymbol{\beta}_I(\hat{\mathbf{d}})=+\boldsymbol{\beta}_I(-\hat{\mathbf{d}}) 
    \end{cases} \quad .
\end{equation}
It is then straightforward to prove by induction that the product $R_\pm=\prod_{k} \rho_\pm(k)$ of an arbitrary set of projectors $\rho_\pm(k)$ is a matrix $R_\pm=B^0(\{\pm\hat{\mathbf{d}}(k)\})\sigma_0+\mathbf{B}(\{\pm\hat{\mathbf{d}}(k)\})\cdot \boldsymbol{\sigma}$ whose $B$-coefficients are functions of the entire set $\{ \hat{\mathbf{d}}(k) \}$ of unit vectors and satisfy the parity relations Eq.(\ref{Eq: parity relations}) in terms of $\{ \hat{\mathbf{d}}(k) \} \leftrightarrow \{ -\hat{\mathbf{d}}(k) \}$, implying that
\begin{eqnarray}
    \text{tr}\prod_{k\in BZ}
     \rho_\pm(k) &=& B^0_R (\{ \pm \hat{\mathbf{d}}(k) \}) +i B^0_I (\{ \pm \hat{\mathbf{d}}(k) \}) = B^0_R (\{  \hat{\mathbf{d}}(k) \}) \pm i B^0_I (\{ \hat{\mathbf{d}}(k) \}) \label{Eq: tr prod rho(k)} \quad .
\end{eqnarray} 
Thus, in view of Eqs.(\ref{Eq: Berry phase explicit gauge ind.}) and (\ref{Eq: tr prod rho(k)}), the relation $\varphi_{B-}+\varphi_{B+}=0 \mod 2\pi$ holds also in the discrete formulation. 

Finally, if charge conjugation symmetry is present then the following relations hold true 
\begin{eqnarray}
    \hat{d}_{x}(k)&=&\hat{d}_{x}(-k) \\
    \hat{d}_{y,z}(k)&=&-\hat{d}_{y,z}(-k)
\end{eqnarray}
and one can prove that $\varphi_{B-}=\varphi_{B+} \mod 2\pi$. Together with the general constraint $\varphi_{B-}=-\varphi_{B+} \mod 2\pi$ this implies that, when the system is invariant under charge conjugation, the Berry phase is constrained to be either $0$ or $\pi$ even in the discrete formulation.

\subsection{The space-time scaling limit}
We can now evaluate the time dependent discrete Berry phase $\varphi_{B-}$ in the space-time scaling limit (STSL). 
Given that $\xi_-(k,t,L)=\langle u_-(k+\delta k,t)| u_-(k,t) \rangle$ and $| u_-(k,t) \rangle    = \exp[-i \left[\mathbf{d}^f (k) t/\hbar \right] \cdot \boldsymbol{\sigma}]  | u_-^i  (k) \rangle $   we have to evaluate 
\begin{equation}
    e^{i \left[ \mathbf{d}^f (k+ \delta k) t/\hbar \right] \cdot \boldsymbol{\sigma} }= \cos \big| \mathbf{d}^f (k+\delta k ) t/\hbar \big| + i \big[ \hat{\mathbf{d}}^f (k+\delta k) \cdot \boldsymbol{\sigma} \big] \sin \big| \mathbf{d}^f (k+\delta k) t/\hbar \big| \label{exp(k+dk)}
\end{equation}
to order $L^{-1}$, where $t\sim L$ while $\eta=2\pi t/L$ is fixed. Let us first  evaluate the modulus. To make the notation lighter, we suppress an overall factor $\hbar^{-1}$ and the dependence on $k$ in the the Taylor expansion $\mathbf{d}^f(k+\delta k)=\mathbf{d}^f + \partial_k \mathbf{d}^f \delta k + \frac{1}{2}\partial^2_k \mathbf{d}^f (\delta k)^2 + O(\delta k)^3$. Thus we get
\begin{eqnarray}
    \big| \mathbf{d}^f (k+\delta k ) t \big|&=&\Big|\mathbf{d}^f t+ \partial_k \mathbf{d}^f  \eta + \frac{1}{2} \delta k \, \partial_k^2 \mathbf{d}^f \eta + O(L^{-2}) \Big|  \nonumber \\
    &=& |\mathbf{d}^f t| \sqrt{1+ 2\frac{\mathbf{d}^f\cdot \partial_k \mathbf{d}^f}{|\mathbf{d}^f|^2}  \delta k + \left(\frac{|\partial_k \mathbf{d}^f|^2}{|\mathbf{d}^f|^2} + \frac{\mathbf{d}^f \cdot \partial_k^2 \mathbf{d}^f}{|\mathbf{d}^f|^2} \right) \delta k^2  +O({L}^{-3})} \nonumber  \\
    &=& |\mathbf{d}^f t| \sqrt{1+ 2\frac{\mathbf{d}^f\cdot \partial_k \mathbf{d}^f}{|\mathbf{d}^f|^2}  \delta k +  \frac{\partial_k(\mathbf{d}^f \cdot \partial_k \mathbf{d}^f)}{|\mathbf{d}^f|^2}  \delta k^2 +O({L}^{-3})} \nonumber   \\
    &=& |\mathbf{d}^f t| \left\{ 1+  \frac{\mathbf{d}^f\cdot \partial_k \mathbf{d}^f}{|\mathbf{d}^f|^2}  \delta k + \left[ \frac{1}{2} \frac{\partial_k(\mathbf{d}^f \cdot \partial_k \mathbf{d}^f)}{|\mathbf{d}^f|^2}    - \left(\frac{\mathbf{d}^f\cdot \partial_k \mathbf{d}^f}{|\mathbf{d}^f|^2}  \right)^2 \right] \delta k^2  +O({L}^{-3}) \right\}   \nonumber \\
    &=&  d^f t \left\{ 1+  \frac{ \partial_k d^f}{d^f}  \delta k + \frac{1}{2} \left[  \frac{d^f \partial_k^2 d^f - (\partial_k d^f)^2}{(d^f)^2}     \right] \delta k^2  +O({L}^{-3}) \right\} \nonumber    \\
    &=&  d^f t  + \partial_k d^f \eta + \frac{1}{2} \delta k \,  d_2^f   \eta   +O({L}^{-2})    
\end{eqnarray}
where $d^f=|\mathbf{d}^f|$ and
\begin{equation}
     d_2^f=\partial_k^2 d^f - \frac{(\partial_k d^f)^2}{d^f} \quad .
\end{equation}
 Then the unit vector $\hat{\mathbf{d}}^f(k+\delta k)$  appearing in Eq.(\ref{exp(k+dk)}) is given by
\begin{eqnarray}
\hat{\mathbf{d}}^f (k+\delta k )       
&=& \frac{\mathbf{d}^f t}{d^f t \left\{ 1+  \frac{ \partial_k d^f}{d^f}  \delta k + \frac{1}{2} \left[  \frac{d^f \partial_k^2 d^f - (\partial_k d^f)^2}{(d^f)^2}     \right] \delta k^2  +O({L}^{-3}) \right\}} \nonumber \\
    && + \frac{\partial_k \mathbf{d}^f  \eta}{d^f t \left\{ 1+  \frac{ \partial_k d^f}{d^f}  \delta k + \frac{1}{2} \left[  \frac{d^f \partial_k^2 d^f - (\partial_k d^f)^2}{(d^f)^2}     \right] \delta k^2  +O({L}^{-3}) \right\}} + O({L}^{-2}) \nonumber \\  
&=&  \frac{\mathbf{d}^f t}{d^f t}  \left\{ 1-  \frac{ \partial_k d^f}{d^f}  \delta k  \right\} + \frac{\partial_k \mathbf{d}^f  \eta}{d^f t } + O({L}^{-2}) \,
 = \hat{\mathbf{d}}^f   + \partial_k \hat{\mathbf{d}}^f   \delta k  + O({L}^{-2})  
\end{eqnarray}
whence one obtains
\begin{eqnarray}
    e^{i \left[ \mathbf{d}^f (k+ \delta k) t/\hbar \right] \cdot \boldsymbol{\sigma} } 
    &=& \cos\big[ (d^f t  + \partial_k d^f \eta + \frac{1}{2} \delta k d_2^f  \eta)/\hbar\big]   + i  \big[(\hat{\mathbf{d}}^f   + \partial_k \hat{\mathbf{d}}^f   \delta k) \cdot \boldsymbol{\sigma}\big] \sin\big[ (d^f t  + \partial_k d^f \eta + \frac{1}{2} \delta k d_2^f  \eta)/\hbar \big]  +O({L}^{-2}) \nonumber \\
    &=& e^{i [ (d^f t  + \partial_k d^f \eta + \frac{1}{2} \delta k d_2^f  \eta)/\hbar] \hat{\mathbf{d}}^f  \cdot \boldsymbol{\sigma}} + i \delta k \sin\big[ (d^f t  + \partial_k d^f \eta)/\hbar \big]    [\partial_k \hat{\mathbf{d}}^f  \cdot \boldsymbol{\sigma} ]   +O({L}^{-2})
\end{eqnarray}
and
\begin{eqnarray}
    e^{i \left[ \mathbf{d}^f (k+ \delta k) t/\hbar \right] \cdot \boldsymbol{\sigma} } e^{-i \left[\mathbf{d}^f(k)  t/\hbar \right]\cdot \boldsymbol{\sigma}} 
    &=& e^{i \left[(   \partial_k d^f \eta + \frac{1}{2} \delta k d_2^f  \eta)/\hbar\right] \hat{\mathbf{d}}^f  \cdot \boldsymbol{\sigma}}   + i \delta k \sin\big[ (d^f t  + \partial_k d^f \eta )/\hbar \big]   [ \partial_k \hat{\mathbf{d}}^f  \cdot \boldsymbol{\sigma} ]  e^{-i\left[ \mathbf{d}^f t/\hbar\right] \cdot \boldsymbol{\sigma}}  +O({L}^{-2}) \nonumber \\
    &=& \cos\big[ (   \partial_k d^f \eta + \frac{1}{2} \delta k d_2^f  \eta ) /\hbar \big]    +i\sin\big[ (  \partial_k d^f \eta + \frac{1}{2} \delta k d_2^f  \eta ) / \hbar \big] [\hat{\mathbf{d}}^f  \cdot \boldsymbol{\sigma}] \nonumber \\
    && + i \delta k \sin\left[ (d^f t  + \partial_k d^f \eta)/\hbar \right]  \cos\left[ d^f t/\hbar \right]  [ \partial_k \hat{\mathbf{d}}^f  \cdot \boldsymbol{\sigma} ]  \nonumber \\
    &&+ i \delta k \sin\left[ (d^f t  + \partial_k d^f \eta)/\hbar \right]   \sin\left[ d^f t /\hbar \right] [( \partial_k \hat{\mathbf{d}}^f  \times \hat{\mathbf{d}}^f )  \cdot \boldsymbol{\sigma}]  +O({L}^{-2}) \nonumber \\
    &=& \cos\big[   \partial_k d^f \eta/\hbar  \big]   - \frac{1}{2}  \sin\big[   \partial_k d^f \eta/\hbar  \big] [\delta k d_2^f \eta/\hbar]  \nonumber \\
    && +i\sin\left[   \partial_k d^f \eta/\hbar  \right] [\hat{\mathbf{d}}^f  \cdot \boldsymbol{\sigma}] + \frac{i}{2}  \cos\left[   \partial_k d^f \eta/\hbar  \right]  \big[ \delta k d_2^f  \eta/\hbar \big] [\hat{\mathbf{d}}^f  \cdot \boldsymbol{\sigma}]\nonumber \\
    && + i \delta k \sin\left[ ( d^f t  + \partial_k d^f \eta )/\hbar\right]  \cos\left[ d^f t/\hbar \right]   [ \partial_k \hat{\mathbf{d}}^f  \cdot \boldsymbol{\sigma} ]  \nonumber \\
    &&+ i \delta k \sin\left[ (d^f t  + \partial_k d^f \eta)/\hbar \right]   \sin\left[ d^f t/\hbar \right] [( \partial_k \hat{\mathbf{d}}^f  \times \hat{\mathbf{d}}^f )  \cdot \boldsymbol{\sigma} ] +O({L}^{-2}) \nonumber \\
    &=& e^{i [ \partial_k d^f \eta/\hbar ] [\hat{\mathbf{d}}^f  \cdot \boldsymbol{\sigma}]}   - \frac{\delta k}{2}   \sin[ \partial_k d^f \eta/\hbar] [d_2^f \eta/\hbar]    + i \frac{\delta k}{2}   \cos [ \partial_k d^f \eta/\hbar ] [d_2^f \eta/\hbar]  [\hat{\mathbf{d}}^f  \cdot \boldsymbol{\sigma}]  \nonumber \\
    && + i \delta k \cos\left[ \partial_k d^f \eta/\hbar \right] \left\{ \sin\left[ d^f t/\hbar  \right]  \cos\left[ d^f t /\hbar\right] [ \partial_k \hat{\mathbf{d}}^f  \cdot \boldsymbol{\sigma} ]+\sin^2\left[ d^f t /\hbar \right]  [ (\partial_k \hat{\mathbf{d}}^f  \times \hat{\mathbf{d}}^f )\cdot \boldsymbol{\sigma} ] \right\}   \nonumber \\
    && + i \delta k \sin\left[ \partial_k d^f \eta/\hbar \right] \left\{ \cos^2\left[ d^f t/\hbar \right] [ \partial_k \hat{\mathbf{d}}^f  \cdot \boldsymbol{\sigma} ]+\cos\left[ d^f t/\hbar \right] \sin\left[ d^f t/\hbar  \right]  [( \partial_k \hat{\mathbf{d}}^f  \times \hat{\mathbf{d}}^f )\cdot \boldsymbol{\sigma} ] \right\} \nonumber \\
    &&+O({L}^{-2}) \quad .
\end{eqnarray}
Thus, by recalling the definition $\xi_-(k,t,L)=\langle u_-(k+\delta k,t)| u_-(k,t) \rangle=\langle u_-^i(k+\delta k)| e^{i \left[ \mathbf{d}^f (k+ \delta k) t/\hbar \right] \cdot \boldsymbol{\sigma} } e^{-i \left[\mathbf{d}^f(k)  t/\hbar \right]\cdot \boldsymbol{\sigma}}  | u_-^i(k) \rangle$ and by Taylor expanding $\langle u_-^i(k+\delta k)|=\langle u_-^i|+\delta k \langle \partial_k u_-^i| +O(\delta k)^2 $, we finally arrive at 
\begin{eqnarray}
    \xi_-(k,t,{L})&=& \langle u_-^i| e^{i [ \partial_k d^f \eta/\hbar ] [\hat{\mathbf{d}}^f  \cdot \boldsymbol{\sigma}]} | u_-^i \rangle   + \delta k \langle \partial_k u_-^i| e^{i [ \partial_k d^f \eta/\hbar ] [\hat{\mathbf{d}}^f  \cdot \boldsymbol{\sigma}]} | u_-^i \rangle \nonumber \\
    && - \frac{\delta k}{2}  \sin[ \partial_k d^f \eta/\hbar ] [d_2^f \eta/\hbar]  \langle u_-^i | u_-^i \rangle   + i \frac{\delta k}{2}  \cos [ \partial_k d^f \eta/\hbar ] [d_2^f \eta/\hbar] \langle u_-^i| [\hat{\mathbf{d}}^f  \cdot \boldsymbol{\sigma}] | u_-^i\rangle \nonumber \\
    && + i \delta k \cos\left[ \partial_k d^f \eta/\hbar \right] \langle u_-^i | \Big\{ \frac{1}{2}\sin\left[ 2 d^f t/\hbar  \right] [ \partial_k \hat{\mathbf{d}}^f  \cdot \boldsymbol{\sigma} ]+\frac{1}{2}\big(1-\cos\left[2 d^f t /\hbar  \right] \big) [ (\partial_k \hat{\mathbf{d}}^f  \times \hat{\mathbf{d}}^f )\cdot \boldsymbol{\sigma} ]\Big\} |u_-^i \rangle  \nonumber \\
    && + i \delta k \sin\left[ \partial_k d^f \eta/\hbar \right]  \langle u_-^i |  \Big\{ \frac{1}{2}\big(1+\cos\left[2 d^f t /\hbar  \right] \big) [ \partial_k \hat{\mathbf{d}}^f  \cdot \boldsymbol{\sigma} ]+ \frac{1}{2} \sin\left[2 d^f t/\hbar  \right]  [ (\partial_k \hat{\mathbf{d}}^f  \times \hat{\mathbf{d}}^f )\cdot \boldsymbol{\sigma}] \Big\} |u_-^i \rangle  +O({L}^{-2})   \nonumber \\
    &=& \xi_-^{(0)}(k,\eta) +\delta k \, \xi_-^{(1)}(k,\eta,t) +O({L}^{-2}) \label{eq_xi_STSL}
\end{eqnarray}
where
\begin{eqnarray}
    \xi_-^{(0)}(k,\eta)&=&\langle u_-^i| e^{i [ \partial_k d^f \eta/\hbar ] [\hat{\mathbf{d}}^f  \cdot \boldsymbol{\sigma}]} | u_-^i \rangle  
  = \cos[\partial_k d^f \eta/\hbar] + i \sin[\partial_k d^f \eta/\hbar] \langle u^i | [\hat{\mathbf{d}}^f  \cdot \boldsymbol{\sigma}] |u^i\rangle \nonumber \\
    &=& \cos[\partial_k d^f \eta/\hbar] + i [\hat{\mathbf{d}}^f \cdot \hat{\mathbf{d}}^i ] \sin[\partial_k d^f \eta/\hbar]  \label{xi(0)}
\end{eqnarray}

\vspace{5pt}
\noindent while $\xi_-^{(1)}(k,\eta,t)=\xi_-^{(1,A)}(k,\eta)+\xi_-^{(1,B)}(k,\eta,t)$ and 
\begin{eqnarray}
  && \hspace{-25pt} \xi_-^{(1,A)}(k,\eta)= - \frac{1}{2}  \sin[ \partial_k d^f \eta/\hbar ] [d_2^f \eta/\hbar]    -   \sin[ \partial_k d^f \eta/\hbar ] \text{Im}\{\langle \partial_k u_-^i|  [\hat{\mathbf{d}}^f  \cdot \boldsymbol{\sigma}] | u_-^i \rangle \} \nonumber \\
   && \hspace{25pt} + \frac{i}{2}  \cos [ \partial_k d^f \eta/\hbar ] [d_2^f \eta/\hbar] \langle u_-^i| [\hat{\mathbf{d}}^f  \cdot \boldsymbol{\sigma}] | u_-^i\rangle + i \sin[ \partial_k d^f \eta/\hbar ] \text{Re}\{\langle \partial_k u_-^i|  [\hat{\mathbf{d}}^f  \cdot \boldsymbol{\sigma}] | u_-^i \rangle\} \nonumber \\
   && \hspace{25pt}  + i  \cos [ \partial_k d^f \eta/\hbar ] \langle  u_-^i | i \partial_k | u_-^i \rangle  +\frac{i}{2}  \cos\left[ \partial_k d^f \eta/\hbar \right] \langle u_-^i |  [ (\partial_k \hat{\mathbf{d}}^f  \times \hat{\mathbf{d}}^f )\cdot \boldsymbol{\sigma} ] |u_-^i \rangle  \nonumber \\
   && \hspace{25pt} + \frac{i}{2}  \sin\left[ \partial_k d^f \eta/\hbar \right]  \langle u_-^i |   [ \partial_k \hat{\mathbf{d}}^f  \cdot \boldsymbol{\sigma} ]  |u_-^i \rangle    \label{Eq:xi(1A)}
\end{eqnarray}
\begin{eqnarray}
   \xi_-^{(1,B)}(k,\eta,t)&=&
    \frac{i}{2}   \sin\left[ 2 d^f t/\hbar  \right] \langle u_-^i | \Big\{ \cos\left[ \partial_k d^f \eta/\hbar \right]  [ \partial_k \hat{\mathbf{d}}^f  \cdot \boldsymbol{\sigma} ] + \sin\left[ \partial_k d^f \eta/\hbar \right]  [ (\partial_k \hat{\mathbf{d}}^f  \times \hat{\mathbf{d}}^f )\cdot \boldsymbol{\sigma}]  \Big\} |u_-^i \rangle  \nonumber \\
    &+&\frac{i}{2} \cos\left[2 d^f t /\hbar  \right]  \langle u_-^i \Big\{ - \cos\left[ \partial_k d^f \eta/\hbar \right]  [ (\partial_k \hat{\mathbf{d}}^f  \times \hat{\mathbf{d}}^f )\cdot \boldsymbol{\sigma} ] + \sin\left[ \partial_k d^f \eta/\hbar \right]  [ \partial_k \hat{\mathbf{d}}^f  \cdot \boldsymbol{\sigma} ]\Big\} |u_-^i \rangle \quad .  \label{Eq:xi(1B)}
\end{eqnarray}
Notice that, since by definition of $\xi_-$ its gauge dependence  amounts to corrections of order $\delta k$,  the zeroth order contribution $\xi_-^{(0)}$ turns out to be {\it gauge invariant}. 
Moreover, the modulus of $\xi_-$ is given by
\begin{eqnarray}
    |\xi_-|&=&\sqrt{|\xi_-^{(0)}|^2 + 2\delta k \,\text{Re} \left\{ \xi_-^{(0)} \xi_-^{(1)} \right\} + O({L}^{-2})}  
    = |\xi_-^{(0)}| +\delta k \,\text{Re} \left\{ \frac{\xi_-^{(0)} \xi_-^{(1)}}{|\xi_-^{(0)}|} \right\}+ O({L}^{-2})
\end{eqnarray}
while the argument equals 
\begin{eqnarray}
    \text{arg} \left\{\xi_- \right\} &=&\text{Im} \left\{ \ln \left[  \xi_-^{(0)} + \delta k \xi_-^{(1)}   + O({L}^{-2}) \right] \right\}  \, \,
    =\text{arg} \left\{\xi_-^{(0)} \right\} + \delta k \, \text{Im} \left\{  \frac{(\xi_-^{(0)})^*\xi_-^{(1)}}{|\xi_-^{(0)}|^2}    \right\} + O({L}^{-2})
\end{eqnarray}
The Berry phase thus acquires the form  
\begin{eqnarray}
        \varphi_{B-}=\sum_{k \in BZ} \arg \xi_-   = \sum_{k \in BZ} \arg \left\{\xi_-^{(0)} \right\} +  \sum_{k \in BZ} \delta k \, \text{Im} \left\{  \frac{(\xi_-^{(0)})^*\xi_-^{(1)}}{|\xi_-^{(0)}|^2}    \right\} + O({L}^{-1}) = \varphi_{B-}^{(0)} + \varphi_{B-}^{(1)} + O({L}^{-1})
\end{eqnarray}
where
\begin{eqnarray}
    \varphi_{B-}^{(0)}&=&\sum_{k \in BZ} \arg \left\{\xi_-^{(0)} \right\} \approx \frac{L}{2\pi} \int_{-\pi}^\pi dk \, \arg \left\{\xi_-^{(0)} \right\} \label{eq_varphi_0}
\end{eqnarray}
while
\begin{equation}
    \varphi_{B-}^{(1)}=\sum_{k \in BZ} \delta k \, \text{Im} \left\{  \frac{(\xi_-^{(0)})^*\xi_-^{(1)}}{|\xi_-^{(0)}|^2}    \right\} \approx \int_{-\pi}^\pi dk \, \text{Im} \left\{  \frac{(\xi_-^{(0)})^*\xi_-^{(1,A)}}{|\xi_-^{(0)}|^2}    \right\} \quad . \label{eq_varphi_1}
\end{equation}
In Eq.(\ref{eq_varphi_0}) the summation over $k$ has been replaced by an integral, according to the usual recipe $\sum_{k \in BZ} \rightarrow (L/2\pi) \int_{-\pi}^\pi dk$, since the function $\arg \{\xi_-^{(0)}(k,\eta) \}$ varies smoothly in $k$ with respect to $\delta k$.
The same procedure cannot be straightforwardly applied to the summation in Eq.(\ref{eq_varphi_1}) since $\xi_-^{(1)}(k,\eta,t)$ contains terms proportional to $\sin\left[2 d^f(k) t/\hbar\right]$ and $\cos\left[2 d^f(k) t/\hbar\right]$, which are rapidly oscillatory functions of $k$ for $t\sim L$ [see the contribution $\xi_-^{(1,B)}(k,\eta,t)$ in Eq.(\ref{Eq:xi(1B)})].
Nonetheless, because of this highly oscillatory behaviour, the summation of such terms is negligible in the STSL and $\varphi_{B-}^{(1)}$ in Eq.(\ref{eq_varphi_1}) is eventually recast into an integral involving the contribution $\xi_-^{(1,A)}(k,\eta)$ only.\\

\subsection{Berry phase and current operator}
We start this Section by showing that, for $\eta\rightarrow 0$, which corresponds to the usual thermodynamic limit, one recovers the known relation $\partial_t \varphi_B(t)/2\pi =J^f(t)$ \cite{Cooper_PRL_2018_SM}.
Indeed, when $\eta\rightarrow 0$, Eq.(\ref{eq_xi_STSL}) reduces to
\begin{eqnarray}
    \xi_-(k,t,{L})&=& 1 + i \delta k (\partial_k d^f t/\hbar ) \langle u_-^i |  [  \hat{\mathbf{d}}^f  \cdot \boldsymbol{\sigma} ]  |u_-^i \rangle  +
   i \delta k \langle u^i |i\partial_k | u^i\rangle  \nonumber \\
    && + i \delta k  \langle u_-^i | \left\{ \frac{1}{2}\sin\left[ 2 d^f t/\hbar  \right] [ \partial_k \hat{\mathbf{d}}^f  \cdot \boldsymbol{\sigma} ]+\frac{1}{2} \big(1-\cos\left[2 d^f t /\hbar  \right] \big) [ (\partial_k \hat{\mathbf{d}}^f  \times \hat{\mathbf{d}}^f )\cdot \boldsymbol{\sigma} ] \right\} |u_-^i \rangle  
 +O({L}^{-2})
\end{eqnarray}
so that
\begin{eqnarray}
    \text{arg} \{ \xi_-(k,t,{L})\} &=&  \delta k \langle u^i |i\partial_k | u^i\rangle + \delta k (\partial_k d^f t/\hbar ) \langle u_-^i |  [  \hat{\mathbf{d}}^f  \cdot \boldsymbol{\sigma} ]  |u_-^i \rangle  
  \nonumber \\
    && \hspace{-15pt} +\delta k  \langle u_-^i | \left\{ \frac{1}{2}\sin\left[ 2 d^f t/\hbar  \right] [ \partial_k \hat{\mathbf{d}}^f  \cdot \boldsymbol{\sigma} ]+\frac{1}{2} \big(1-\cos\left[2 d^f t /\hbar  \right] \big) [ (\partial_k \hat{\mathbf{d}}^f  \times \hat{\mathbf{d}}^f )\cdot \boldsymbol{\sigma} ]  \right\} |u_-^i \rangle  +
 O({L}^{-2})
\end{eqnarray}
and 
\begin{eqnarray}
    \frac{\varphi_{B-}(t)}{2\pi}&=& \frac{1}{2\pi}\sum_k \text{arg} \{ \xi_-(k,t,{L})\} \, 
     \approx 
  \frac{1}{2\pi} \int_{-\pi}^{\pi} dk A_{B-}^i(k) + \frac{1}{2\pi}\int_{-\pi}^{\pi} dk  [\partial_k d^f/\hbar] \langle u_-^i |  [  \hat{\mathbf{d}}^f  \cdot \boldsymbol{\sigma} ]  |u_-^i \rangle \, t  \nonumber \\
    && +  \frac{1}{2\pi} \int_{-\pi}^{\pi} dk  \langle u_-^i | \left\{ \frac{1}{2}\sin\left[ 2 d^f t/\hbar  \right] [ \partial_k \hat{\mathbf{d}}^f  \cdot \boldsymbol{\sigma} ]+\frac{1}{2} \big(1-\cos\left[2 d^f t /\hbar  \right] \big) [ (\partial_k \hat{\mathbf{d}}^f  \times \hat{\mathbf{d}}^f )\cdot \boldsymbol{\sigma} ]  \right\} |u_-^i \rangle  \nonumber \\
    &=&  \frac{\varphi_{B-}(0)}{2\pi}+ J^f_{DC} \, t  + \int_{0}^{t} dt' J^f_{AC}(t')
\end{eqnarray}
where $A_{B-}^i(k)$ is the initial Berry connection, while $J^f_{DC}$ and $J^f_{AC}(t)$ are the  integral over the Brillouin zone of the  expectation values of  the operators $\mathcal{J}^f_{DC}(k)=\hbar^{-1} \partial_k d^f (\hat{\mathbf{d}^f} \cdot \boldsymbol{\sigma})$ and $\mathcal{J}^f_{AC}(k,t)=\hbar^{-1} d^f  \{ \cos[2d^f t/\hbar] [\partial_k\hat{\mathbf{d}^f} \cdot \boldsymbol{\sigma}] +  \sin[2d^f t/\hbar] [(\hat{\mathbf{d}}^f\times\partial_k\hat{\mathbf{d}^f})\cdot  \boldsymbol{\sigma}]  \,   \}$, respectively, which can in turn be interpreted as the  DC/AC term of $k$-component current operator  resulting from its Heisenberg evolution 
\begin{eqnarray}
    \hbar \mathcal{J}^f(k,t)&=&e^{-i  [d^ft/\hbar] \hat{\mathbf{d}}^f\cdot \boldsymbol{\sigma}} [\partial_k \mathbf{d}^f \cdot \boldsymbol{\sigma}] e^{i  [d^ft/\hbar] \hat{\mathbf{d}}^f\cdot \boldsymbol{\sigma}}  \nonumber \\
    &=& e^{-i  [d^ft/\hbar] \hat{\mathbf{d}}^f\cdot \boldsymbol{\sigma}} [\partial_k d^f (\hat{\mathbf{d}^f} \cdot \boldsymbol{\sigma}) +d^f (\partial_k\hat{\mathbf{d}^f} \cdot \boldsymbol{\sigma})] e^{i  [d^ft/\hbar] \hat{\mathbf{d}}^f\cdot \boldsymbol{\sigma}}  \nonumber \\
    &=& \partial_k d^f (\hat{\mathbf{d}^f} \cdot \boldsymbol{\sigma})+d^f \Big( \cos[d^f t/\hbar] -i \sin[d^f t/\hbar] [\hat{\mathbf{d}}^f\cdot \boldsymbol{\sigma}] \Big) [ \partial_k\hat{\mathbf{d}^f} \cdot \boldsymbol{\sigma}] \Big( \cos[d^f t/\hbar] +i \sin[d^f t/\hbar] [\hat{\mathbf{d}}^f\cdot \boldsymbol{\sigma}] \Big) \nonumber \\
    &=& \partial_k d^f (\hat{\mathbf{d}^f} \cdot \boldsymbol{\sigma})  +d^f \Big( \cos^2[d^f t/\hbar] [\partial_k\hat{\mathbf{d}^f} \cdot \boldsymbol{\sigma}] -i  \sin[d^f t/\hbar]\cos[d^f t/\hbar] [\hat{\mathbf{d}}^f\cdot \boldsymbol{\sigma}] [\partial_k\hat{\mathbf{d}^f} \cdot \boldsymbol{\sigma}] \nonumber \\
    &&\qquad +i  \sin[d^f t/\hbar]\cos[d^f t/\hbar] [\partial_k\hat{\mathbf{d}^f} \cdot \boldsymbol{\sigma}] [\hat{\mathbf{d}}^f\cdot \boldsymbol{\sigma}] +\sin^2[d^f t/\hbar] [\hat{\mathbf{d}}^f\cdot \boldsymbol{\sigma}][\partial_k\hat{\mathbf{d}^f} \cdot \boldsymbol{\sigma}][\hat{\mathbf{d}}^f\cdot \boldsymbol{\sigma}] \Big)  \nonumber\\
    &=& \partial_k d^f (\hat{\mathbf{d}^f} \cdot \boldsymbol{\sigma})  +d^f \Big( \cos^2[d^f t/\hbar] [\partial_k\hat{\mathbf{d}^f} \cdot \boldsymbol{\sigma}] +2  \sin[d^f t/\hbar]\cos[d^f t/\hbar] [(\hat{\mathbf{d}}^f\times\partial_k\hat{\mathbf{d}^f})\cdot  \boldsymbol{\sigma}]  \nonumber \\
    &&\qquad -\sin^2[d^f t/\hbar] [\hat{\mathbf{d}}^f\times(\partial_k\hat{\mathbf{d}^f}\times \hat{\mathbf{d}}^f)] \cdot \boldsymbol{\sigma} \Big)  \nonumber\\
    &=& \partial_k d^f (\hat{\mathbf{d}^f} \cdot \boldsymbol{\sigma})  +d^f \Big( \cos^2[d^f t/\hbar] [\partial_k\hat{\mathbf{d}^f} \cdot \boldsymbol{\sigma}] +2  \sin[d^f t/\hbar]\cos[d^f t/\hbar] [(\hat{\mathbf{d}}^f\times\partial_k\hat{\mathbf{d}^f})\cdot  \boldsymbol{\sigma}]   -\sin^2[d^f t/\hbar] [\partial_k \hat{\mathbf{d}}^f \cdot \boldsymbol{\sigma}] \Big)  \nonumber\\
    &=& \partial_k d^f (\hat{\mathbf{d}^f} \cdot \boldsymbol{\sigma})  +d^f \Big( \cos[2d^f t/\hbar] [\partial_k\hat{\mathbf{d}^f} \cdot \boldsymbol{\sigma}] +  \sin[2d^f t/\hbar] [(\hat{\mathbf{d}}^f\times\partial_k\hat{\mathbf{d}^f})\cdot  \boldsymbol{\sigma}]  \,  \Big) 
\end{eqnarray}

Let us now determine the relation between $\varphi_{B-}^{(0)}$ and the current operator in the space-time scaling limit. Starting from  $\xi_-^{(0)}(k,\eta) = \langle u_-^i(k) | e^{i\mathcal{J}^f_{DC}(k) \eta} |u_-^i(k)\rangle$ we get 
\begin{eqnarray}
    \frac{1}{i}\partial_\eta \xi_-^{(0)}(k,\eta)&=& \langle u_-^i(k) | e^{i\mathcal{J}_{DC}^f(k) \eta} \mathcal{J}_{DC}^f(k) \sum_{s=\pm}|u_s^i(k)\rangle \langle u_s^i(k)|   u_-^i(k)\rangle= \nonumber  \\
    &=&  \xi_-^{(0)}(k,\eta)  \langle u_-^i(k)| \mathcal{J}_{DC}^f(k) |u_-^i(k)\rangle + \chi_-^{(0)}(k,\eta) \langle u_+^i(k)| \mathcal{J}_{DC}^f(k) |u_-^i(k)\rangle 
\end{eqnarray}
where $ \chi_-^{(0)}(k,\eta)= \langle u_-^i(k) | e^{i\mathcal{J}_{DC}^f(k) \eta} |u_+^i(k)\rangle$. Then, in view of the definition
\begin{eqnarray}
\varphi_{B-}^{(0)}(\eta)&=&\frac{L}{2\pi} \int_{-\pi}^\pi dk \, \arg \xi_-^{(0)}(k,\eta) =
 \frac{L}{2\pi} \int_{-\pi}^\pi dk \, \text{Im} \ln \xi_-^{(0)}(k,\eta) 
\end{eqnarray}
we arrive at
\begin{eqnarray}
\frac{d}{d\eta} \varphi_{B-}^{(0)}(\eta)&=& \frac{L}{2\pi} \int_{-\pi}^\pi dk \, \text{Im} \frac{d}{d\eta} \ln \xi_-^{(0)}(k,\eta) \nonumber \\
&=& \frac{L}{2\pi} \int_{-\pi}^\pi dk \, \text{Im} \frac{i \xi_-^{(0)}(k,\eta)  \langle u_-^i(k)| \mathcal{J}_{DC}^f(k) |u_-^i(k)\rangle + i \chi_-^{(0)}(k,\eta) \langle u_+^i(k)| \mathcal{J}_{DC}^f(k) |u_-^i(k)\rangle}{\xi_-^{(0)}(k,\eta)} \nonumber \\
&=& \frac{L}{2\pi} \int_{-\pi}^\pi dk \,\langle u_-^i(k)| \mathcal{J}_{DC}^f(k) |u_-^i(k)\rangle +\frac{L}{2\pi} \int_{-\pi}^\pi dk \, \text{Re} \left\{ \frac{ \chi_-^{(0)}(k,\eta)}{\xi_-^{(0)}(k,\eta)}  \langle u_+^i(k)| \mathcal{J}_{DC}^f(k) |u_-^i(k)\rangle \right\}
\end{eqnarray}


\begin{thebibliography}{51}
\bibitem [{foo()}]{footnote_Wannier}%
  \BibitemOpen
  \href@noop {} {\ }\bibinfo {note} {On general ground, the correlation length
  is expected to grow linearly in time after the quench. In the present
  setting, the correlation length can also be identified with the localization
  length of the maximally localized single particle Wannier functions and the
  velocity of the correlation spreading is dictated by $v^f(k)$. Thus, the
  localization length of the Wannier functions grows linearly in time until the
  space-time scaling limit is reached, when it has become comparable to the
  system size.}\BibitemShut {Stop}%
\end{thebibliography}
\end{document}